\documentclass[aps,pre,twocolumn,amsmath,amssymb]{revtex4-2}

\usepackage{mathrsfs}
\usepackage{graphicx}
\usepackage{adjustbox}
\usepackage{longtable}
\usepackage{booktabs}
\usepackage{bm}
\usepackage{threeparttable}
\usepackage{amsmath,amssymb}
\usepackage{color}
\usepackage{epstopdf}
\usepackage{microtype}
\usepackage{tikz}
\usepackage{mathtools}
\usepackage[colorlinks=true,linkcolor=blue,anchorcolor=blue, citecolor=blue,urlcolor=blue]{hyperref}

\def\T{{\theta_T}}
\def\Tp#1{{\theta_T^{#1}}}
\def\h{\Delta t}
\def\n{\lambda}
\def\a{\alpha}
\def\b{\beta}
\def\g{\gamma}

\def\th{\theta}

\def\ovl#1{\overline{#1}}

\def\df{{\rm d}}

\def\bk#1{\left\langle#1\right\rangle}
\def\ave#1{\mathbb{E}\left[ #1 \right]}
\def\bbkk#1{\kern 0.1em \left\langle \kern -0.3em \left\langle \kern 0.1em #1 \kern 0.1em \right\rangle \kern -0.3em \right\rangle \kern 0.1em}

\def\vb#1{\bm{#1}}
\def\IntV{\int_{-\infty}^{+\infty}d\vb{v}}
\def\pd#1{\partial_{#1}}
\def\pdu#1#2{\partial_{#1}^{#2}}

\def\kr#1{\delta_{#1}}

\def\krv#1{\delta_{#1}^{\perp}}
\def\ep#1{\varepsilon_{#1}}

\allowdisplaybreaks[4]
\begin{document}

\preprint{APS/123-QED}

\title{Heat and mass transport in a three-dimensional mesoscale odd fluid}

\author{Yuxing Jiao}
\email{jiaoyuxing22@mails.ucas.ac.cn}
\affiliation{Beijing National Laboratory for Condensed Matter Physics and Laboratory of Soft Matter Physics, Institute of Physics, Chinese Academy of Sciences, Beijing 100190, China}\affiliation{School of Physical Sciences, University of Chinese Academy of Sciences, Beijing 100049, China}

\author{Mingcheng Yang}
\email{mcyang@iphy.ac.cn}
\affiliation{Beijing National Laboratory for Condensed Matter Physics and Laboratory of Soft Matter Physics, Institute of Physics, Chinese Academy of Sciences, Beijing 100190, China}\affiliation{School of Physical Sciences, University of Chinese Academy of Sciences, Beijing 100049, China}

\begin{abstract}
Fluids with nonvanishing antisymmetric components of the transport coefficient tensor are named odd fluids. In our previous works, we proposed a mesoscale simulation model for isotropic two-dimensional odd fluids and then extended it to the three-dimensional case to model an anisotropic odd fluid with cylindrical symmetry. The Navier-Stokes equation and the viscosity tensor of this three-dimensional mesoscale odd fluid were derived previously via a kinetic theory. Herein, we derive the heat conduction and self-diffusion equations, along with the corresponding thermal conductivity and self-diffusivity tensors. These theoretical results are validated by simulations. We further investigate heat and mass transport behaviors of three-dimensional odd fluids in a confined geometry through our mesoscale model. In striking contrast to normal isotropic fluids, the steady-state temperature and density distributions of confined odd fluids are significantly deformed by the odd transport coefficients.
\end{abstract}

\maketitle

\section{Introduction}
Odd fluids refer to fluids whose transport coefficient tensors possess nonvanishing antisymmetric components. This requires broken time-reversal and parity symmetries in the microscopic dynamics of odd fluids~\cite{NET}. The nonvanishing antisymmetric parts of transport tensors result in transverse transport phenomena in odd fluids, meaning that a general force applied to a fluid can drive a flux perpendicular to it~\cite{PolyGas,PolyGas3_Dufor,PolyGas4_thermaldiffusion,OddEffects1_HallTransport,OddEffects2_TopoTrans,OddEffects3_TopoWave,OddEffects4_OddMob,OddEffects5_Turb,Banerjee2017-fe,Hargus2021}. Thus, compared to normal fluids, odd fluids exhibit more diverse hydrodynamics and transport phenomena. As a result, odd fluids have become a field that has attracted extensive attention. Studies of odd fluids include electron Hall fluids~\cite{Avron1995,Hoyos2012,Bandurin2019,Holder2019}, polyatomic gases within a magnetic field~\cite{PolyGas,PolyGas2_thCond,PolyGas3_Dufor,PolyGas4_thermaldiffusion}, chiral active fluids~\cite{ActiveColloidal,Vitelli_Fluhydro,Markovich2021,Hargus2021}, and so on. The transverse transport property also reflects in the exotic dynamics of mesoscale objects immersed in odd fluids~\cite{OddEffects2_TopoTrans,Yang_2021,Khain2024}. Therefore, the odd complex fluid (the mixture of the odd fluid and suspended mesoscale objects) becomes a fascinating and promising topic.

The most common simulation approach used in the present studies of odd fluids is limited to the molecular-dynamics-type (MD) method, which is unsuitable and even incapable of simulating large-scale odd fluid systems and odd complex fluids. The reason is that the cross-scale and diverse interactions in odd complex fluids create a computational bottleneck for MD simulations. Over past decades, the similar problem has been addressed in conventional complex fluid simulations through the development of various mesoscopic fluids simulation approaches such as the lattice Boltzmann method~\cite{LBE,Ladd1_1994,Ladd2_1994}, dissipative particle dynamics~\cite{DPD,Hoogerbrugge_1992}, and multi-particle collision dynamics (MPC)~\cite{MPC1,MPC2,MPC_MD,Kapral2008,Gompper_2009}. These mesoscopic models not only correctly describe the hydrodynamic behaviors of fluids, but also extremely improve the efficiency of simulations for the solvent. 

However, a mesoscopic model for odd fluids is still lacking. To overcome this, we recently developed the chiral stochastic rotation dynamics (CSRD)~\cite{CSRD,CSRD_3D} by generalizing the stochastic rotation dynamics (SRD), which is a widely-used version of MPC. In our previous work, we have demonstrated that the CSRD inherits all advantages of the traditional SRD: high efficiency, correct hydrodynamics behavior, natural thermal fluctuation (including fluctuation-dissipation theorem), and easy realization of diverse couplings with immersed objects. 

Utilizing a kinetic theory~\cite{Pooley}, we have obtained the complete hydrodynamic equations and corresponding relations for two-dimensional (2D) CSRD~\cite{CSRD}, including the continuity equation, the Navier-Stokes equation, the heat conduction equation, and the self-diffusion equation, which align with the 2D isotropic odd fluid hydrodynamics. Subsequently, we extended the 2D-CSRD to the 3D case~\cite{CSRD_3D}. Unlike the 2D case, 3D odd fluids must be anisotropic because of symmetry restrictions. Consequently, the 3D-CSRD was developed for modeling the most common 3D anisotropic odd fluids with $C_\infty$ cylindrical symmetry: rotational symmetry around a fixed axis (hereafter denoted as the $z$-axis), and broken reflection symmetry across planes containing the $z$-axis. In the earlier work~\cite{CSRD_3D}, the continuity equation, the Navier-Stokes equation, and the stress constitutive relation for 3D-CSRD were derived from the same kinetic theory and verified in simulations. To pursue the complete hydrodynamics for the 3D-CSRD, we continue these kinetic derivations to obtain the heat conduction and self-diffusion equations.

\section{The 3D-CSRD model}
The CSRD model describes fluids by a set of $\mathcal{N}$ point particles with mass $m$. The evolution for position $\bm{r}_i$ and velocity $\bm{v}_i$ of the CSRD particle $i$ follows a discrete-time dynamics with a time step $\h$. The dynamics of CSRD fluid consists of two alternate steps: the streaming step and the collision step. First, in the streaming step, the particle moves ballistically:
\begin{equation}\label{strEq}
    \bm{r}_i\left( t+\h \right)=\bm{r}_i(t)+\bm{v}_i(t)\h.
\end{equation}
After this, in the collision step, the space is first divided into cells using a cubic lattice with a size $l$. To preserve the Galilean invariance, the lattice is shifted by a random displacement vector every CSRD step~\cite{MPC_RS}. Particles within the same cell participate in the following collision, in which their velocities relative to the center-of-mass velocity of the cell $\bm{v}_{cm}$ are rotated:
\begin{equation}\label{colEq}
    \bm{v}_i\left( t+\h \right)=\bm{v}_{cm}+\bm{R}\cdot\left(\bm{v}_i(t)-\bm{v}_{cm}\right).
\end{equation}
Here, $\bm{R}=\bm{R}^{(2)}\cdot\bm{R}^{(1)}$ is the rotation matrix which is composed of a stochastic rotation $\bm{R}^{(1)}$ and a deterministic rotation $\bm{R}^{(2)}$. The stochastic rotation is given by $\bm{R}^{(1)}=\bm{R}^{(1)}\left( \bm{n},\omega \right)$, where $\bm{n}$ is a random rotation axis that is uniformly distributed on the surface of a unit sphere and is chosen independently for each cell, and $\alpha$ is a fixed rotation angle. The deterministic rotation operation $\bm{R}^{(2)}$ is a rotation around $z$-axis by a fixed angle $\theta$, i.e., $\bm{R}^{(2)}=\bm{R}^{(2)}\left( \bm{e}_z,\theta \right)$. If we set $\theta=0$, then the CSRD reduces to the traditional SRD. Actually, it is just the additional rotation $\bm{R}^{(2)}$ that breaks time-reversal and parity symmetries and thus enables the 3D-CSRD to describe an anisotropic odd fluid with the $C_\infty$ cylindrical symmetry. One can verify that CSRD conserves the mass exactly and locally conserves the momentum and energy on the cell level, which is the same as in SRD.

\section{Heat conduction equation for the 3D-CSRD fluid}
The hydrodynamic equations for the 3D-CSRD can be derived from a kinetic method~\cite{CSRD_3D}, originally proposed by Pooley and Yeomans for the traditional SRD~\cite{Pooley}. In this section, we utilize this kinetic theory to derive the energy conservation equation for the 3D-CSRD fluid, which will result in the heat conduction equation.

We first provide the basic settings of this kinetic method. The derivations are performed under the molecular-chaos hypothesis, so we only use the single-particle distribution function $f\left( \bm{r},\bm{v} \right)$ with the normalization condition $\int\df\bm{r}\df\bm{v}f=m\mathcal{N}$. The mass density is then given by $\rho\left( \bm{r} \right)=\int\df\bm{v}f$. We then define the space distribution of a given quantity $X=X\left( \bm{r},\bm{v} \right)$ by $\bk{X\left( \bm{r},\bm{v} \right)}\triangleq\frac{1}{\rho}\int\df\bm{v}Xf$. The flow field is the first velocity moment $\bm{u}\left( \bm{r} \right)=\bk{\bm{v}}$. The higher velocity moments are denoted as $M_{\a\b\cdots}\triangleq\bk{\left( v_\a-u_\a \right)\left( v_\b-u_\b \right)\cdots}$. Notice that the second moment is related to the temperature $k_BT$: $\T\triangleq k_BT/m=\frac{1}{3}M_{\a\a}$. We assume the CSRD fluid is near-equilibrium, such that $f\left( \bm{r},\bm{v} \right)$ is regarded as the local-equilibrium distribution and can be expressed in terms of local thermodynamic quantities:
\begin{equation}\label{localEqdistribution}
    f(\bm{r},\bm{v})=\frac{\rho(\bm{r})}{\Tp{3/2}(\bm{r})}g\left( \frac{\bm{v}-\bm{u}(\bm{r})}{\sqrt{\T(\bm{r})}} \right)
\end{equation}
with $g(x)$ being a function of the dimensionless quantity $x$. We focus on the hydrodynamics limit of the CSRD, for which we assume the space distribution of conserved quantities vary slowly on time and space. Hence we consider the $n$th-order space and time derivative to be of order $\mathcal{O}\left( \delta^n \right)$. 

We write down the conservation equations of mass, momentum, and energy by a general form as follows:
\begin{equation}\label{Q}
    \pd{t}\rho_Q+\pd{\a}J_\a^{(Q)}=0,
\end{equation}
where $\rho_Q=\bk{Q}$ is the density of the conserved quantity $Q=Q\left( \bm{v} \right)$ and $J_\a^{(Q)}$ is the corresponding flux. The flux $J_\a^{(Q)}$ can be expressed as~\cite{CSRD_3D}:
\begin{equation}\label{Flux_DtoC}
    J_\a^{(Q)}\left( t \right)=j_\a^{(Q)}\left( t \right)-\frac{\h}{2}\pd{t}j_\a^{(Q)}\left( t \right)+\mathcal{O}\left( \delta^2 \right).
\end{equation}
Here, $j_\a^{(Q)}\left( t \right)$ is the ``discrete flux'', which represents the time average flux during $\left[ t,t+\h \right]$ (i.e., one CSRD step). In Eq.~\eqref{Flux_DtoC}, we have used the average flux in $[t-\h,t+\h]$ to approximate the flux $J_\a^{(Q)}(t)$. The discrete flux $j_\a^{(Q)}$ consists of a \emph{kinetic part} $j_\a^{(Q),\rm{kin}}$ (contributed by the streaming step) and a \emph{collisional part} $j_\a^{(Q),\rm{col}}$ (contributed by the collision step): $j_\a^{(Q)}=j_\a^{(Q),\rm{kin}}+j_\a^{(Q),\rm{col}}$. The discrete flux can be derived by using Eqs.~\eqref{strEq},\eqref{colEq}, and \eqref{localEqdistribution}. In the following, we calculate the kinetic and collisional energy flux respectively. We set $Q=E_k=\frac{1}{2}m\bm{v}^2$ hereafter.

\subsection{The collisional energy flux}
The average energy flux during the collision step can be derived by using the energy conservation in a collision cell. Here, we focus on a cell divided by an $\alpha$-plane (with a normal vector $\bm{e}_\a$). Then the discrete collisional energy flux across this plane $j_\a^{(Q),\rm{col}}$ can be derived by calculating the average energy variation in the upper-half region of the cell. We choose the center of this cell as the origin and denote the position of the $\alpha$-plane as $x_\a=c$. Note that the position $c$ is a random variable, which is uniformly distributed in $\left[ -l/2,l/2 \right]$ because of the random shift operation. Then the discrete collisional energy flux is expressed as (see also Eq.~(A6) in~\cite{CSRD_3D}):
\begin{equation}\label{qCol}
    \begin{aligned}
        j^{(E_k),\rm{col}}&=\ave{\frac{1}{l^2\h}\prod_{\mu\neq\beta}\left( \int_{-l/2}^{l/2}dr_\mu \right)\right.\\
        &\cdot\left.\int_{c}^{l/2}dr_\b \frac{1}{2}\rho\bk{v_\a^cv_\a^c-v_\a v_\a}}.
    \end{aligned}
\end{equation}
Here, $v_\a^c$ is the post-collision velocity of a particle (say the selected particle) with pre-collision velocity $\bm{v}$ and position $\bm{r}$. Using Eq.~\eqref{colEq}, $v_\a^c$ is given by the following ``single particle collision formula'':
\begin{equation}\label{SingleColEq}
    \begin{aligned}
        v_\a^c&=v_\a-L_{\a\b}\left( v_\b-\hat{v}_\b \right),\\
        L_{\a\b}&\triangleq\frac{N-1}{N}\left( \kr{\a\b} - R_{\a\b} \right),
    \end{aligned}
\end{equation}
where $N\geqslant1$ is the particle number in the cell and $\hat{v}_\a$ is the mean velocity of other particles defined by:
\begin{equation}\label{Vcm}
    v_{cm,\a}=\frac{1}{N}v_\a+\frac{N-1}{N}\hat{v}_\a.
\end{equation}
If we assume the particle number in a cell follows a Poisson distribution with expectation $\n$, then the distribution of $N$ reads: $P(N=q)=e^{-\n}\n^{q-1}/(q-1)!,q\geqslant1$. The rotation matrix and its average are given by~\cite{CSRD_3D}:
\begin{equation}\label{Rmat}
    \begin{aligned}
        R_{\a\b}=&\,\cos\omega\cos\theta\kr{\a\b}+\left( 1-\cos\omega \right)\cos\theta n_\a n_\b\\
                &-\sin\omega\cos\theta n_\g \ep{\g\a\b}+\cos\omega\left( 1-\cos\theta \right)\kr{z\a}\kr{z\b}\\
              &+\left( 1-\cos\omega \right)\left( 1-\cos\theta \right)\kr{z\a}n_z n_\b\\
              &+\sin\omega\left( 1-\cos\theta \right)\kr{z\a}n_\g\ep{z\g\b}\\
              &-\cos\omega\sin\theta\ep{z\a\b}-\left( 1-\cos\omega \right)\sin\theta\ep{z\a\g}n_\g n_\b\\
              &+\sin\omega\sin\theta\left( n_\a\kr{z\b}-n_z\kr{\a\b} \right),
    \end{aligned}
\end{equation}
and
\begin{equation}\label{aveR}
    \begin{aligned}
        &\ave{R_{\a\b}}\\
        &=\frac{1}{3}\left( 1+2\cos\omega \right)\left[ \cos\theta\kr{\a\b}+\left( 1-\cos\theta \right)\kr{z\a}\kr{z\b}\right.\\
        &\qquad\qquad\qquad\qquad\left.-\sin\theta\ep{z\a\b} \right].
    \end{aligned}
\end{equation}
Substituting Eqs.~\eqref{SingleColEq} and \eqref{aveR} into the discrete collisional energy flux Eq.~\eqref{qCol} and applying the molecular-chaos hypothesis leads to:
\begin{equation}\label{qCol_fin}
\begin{aligned}
        &j^{(E_k),\rm{col}}\\
        =\,&-\frac{m\n}{12l\h}\ave{\frac{N-1}{N^2}}\\
        &\cdot\left[ 3-\frac{1}{3}\left( 1+2\cos\omega \right)\left( 1+2\cos\theta \right) \right]\pd{\a}\T+\mathcal{O}\left( \delta^2 \right).
\end{aligned}
\end{equation}
Therefore, according to Eq.~\eqref{Flux_DtoC}, the collisional energy flux is $J_\a^{\left( E_k \right),\rm{col}}=j^{(E_k),\rm{col}}+\mathcal{O}\left( \delta^2 \right)$. The collisional energy flux is also the collisional heat flux, which is denoted by $q_\a$ hereafter.   Then, from Eq.~\eqref{qCol_fin}, we obtain the collisional thermal conductivity tensor $\kappa_{\a\b}^{\rm{col}}$ as:
\begin{equation}\label{kappa_col}
    \begin{aligned}
        &q_\a^{\rm{col}}=J_\a^{\left( E_k \right),\rm{col}}=-\kappa_{\a\b}^{\rm{col}}\pd{\b}T,\qquad\kappa_{\a\b}^{\rm{col}}=\kappa^{\rm{col}}\kr{\a\b},\\
        &\kappa^{\rm{col}}=\frac{k_B\n}{12l\h}\ave{\frac{N-1}{N^2}}\\
        &\qquad\cdot\left[ 3-\frac{1}{3}\left( 1+2\cos\omega \right)\left( 1+2\cos\theta \right) \right].
    \end{aligned}
\end{equation}
In the large $\lambda$ limit, we simply replace $N$ by $\lambda$ for an approximation and obtain:
\begin{equation}
    \begin{aligned}
        \kappa^{\rm{col}}=&\,\frac{k_B}{12l\h}\frac{\n-1}{\n}\left[ 3-\frac{1}{3}\left( 1+2\cos\omega \right)\left( 1+2\cos\theta \right) \right].
    \end{aligned}
\end{equation}
Notice that because of the molecular-chaos hypothesis used in our derivation, the odd part of collisional thermal conductivity tensor is vanishing.

\subsection{The kinetic energy flux}
The kinetic discrete flux can be derived by calculating the average flux across a given plane during the streaming step. Without loss of generality, we choose an area $\mathcal{D}=\left\{ \left( x,y,z \right) | \left\lvert x \right\rvert,\left\lvert z \right\rvert\leqslant \frac{a}{2},\,y=0 \right\}$ to calculate the $y$-component of the flux by (the case of other components are similar):
\begin{equation}\label{jy_kin}
    j_{y}^{(Q),\rm{kin}}=\frac{1}{a^2\h}\IntV\int_{A}d\bm{r}Q\left( \bm{v} \right)f/m,
\end{equation}
where
\begin{equation}
    \begin{aligned}
        A=&\left\{ \left( r_x,r_y,r_z \right) | -v_y\h\leqslant r_y\leqslant 0,\left\lvert r_x-\frac{v_x}{v_y}r_y \right\rvert\leqslant \frac{a}{2},\right.\\
        &\left.\left\lvert r_z-\frac{v_z}{v_y}r_y \right\rvert\leqslant \frac{a}{2} \right\}.
    \end{aligned}
\end{equation}
In~\cite{CSRD_3D} we have calculated the kinetic discrete energy flux via this integral to obtain (see also Eq.~(53) in~\cite{CSRD_3D}):
\begin{equation}\label{Kinqa_D}
    j_{\a}^{(E_k),\rm{kin}}=\frac{5}{2}\rho\T u_\a+\frac{1}{2}\rho M_{\b\b\a}-\frac{5}{4}\h\pd{\a}\left(\rho\Tp{2}\right)+\mathcal{O}(\delta^2).
\end{equation}
Substituting Eq.~\eqref{Kinqa_D} into Eq.~\eqref{Flux_DtoC} results in the kinetic energy flux:
\begin{equation}\label{Kinqa_C}
    \begin{aligned}
        J_\a^{\left( E_k \right),\rm{kin}}=&\,\frac{5}{2}\rho\T u_\a+\frac{1}{2}\rho M_{\b\b\a}-\frac{5}{4}\h\pd{\a}\left(\rho\Tp{2}\right)\\
        &-\frac{\h}{2}\rho\T\pd{t}u_\a+\mathcal{O}(\delta^2).
    \end{aligned}
\end{equation}
According to the Navier-Stokes equation $\pd{t}\left( \rho u_\a \right)+\pd{\b}\left( \rho u_\a u_\b \right)=\pd{\b}\sigma_{\a\b}$ and $\sigma_{\a\b}=-\rho\T\kr{\a\b}+\mathcal{O}\left( \delta \right)$, we have the relation 
\begin{equation}\label{NS0order}
    \rho\pd{t}u_\a=-\pd{\a}\left( \rho\T \right)+\mathcal{O}\left( \delta^2 \right).
\end{equation}
Under this relation, the flux $J_\a^{\left( E_k \right),\rm{kin}}$ becomes
\begin{equation}\label{qKin}
    J_\a^{\left( E_k \right),\rm{kin}}=\frac{5}{2}\rho\T u_\a+\frac{1}{2}\rho M_{\b\b\a}-\frac{5}{4}\rho\T\h\pd{\a}\T.
\end{equation}
In the following, we derive the explicit expression for the third-moment $M_{\b\b\a}$.

\subsubsection{Derivations for the third-moment}
The velocity moments can be solved by a self-consistent equation. We denote the moment at time $t$ by a superscript $t$, i.e., $M_{\a\b\cdots}^t$. The moment $M_{\a\b\cdots}^t$ will be transformed successively by the streaming step and then the collision step, finally becoming $M_{\a\b\cdots}^{t+\h}$. In the steady state, the moment should be stationary: $M_{\a\b\cdots}^t=M_{\a\b\cdots}^{t+\h}=M_{\a\b\cdots}$. Therefore, if we introduce operators $\hat{\mathcal{F}}^s$ and $\hat{\mathcal{F}}^c$ to represent the operations acting on the moment during streaming and collision step respectively, i.e., $M_{\a\b\cdots}^{t+\h}=\hat{\mathcal{F}}^s\circ\hat{\mathcal{F}}^c\left(  M_{\a\b\cdots}^t \right)$. Thus, the following self-consistent equation will be established in the steady state:
\begin{equation}
    M_{\a\b\cdots}=\hat{\mathcal{F}}^s\circ\hat{\mathcal{F}}^c \left( M_{\a\b\cdots} \right).
\end{equation}
Proceeding, we calculate these two transforms on the third-moment.

\emph{Transformation in the streaming step.}---We use superscript $s$ to denote quantities after streaming. According to Eq.~\eqref{strEq}, the single particle distribution function after streaming becomes $f^s\left( \bm{r},\bm{v} \right)=f^t\left( \bm{r}-\bm{v}\h,\bm{v} \right)$. Therefore $M_{\a\b\g}^s=\bk{\left( v_\a-u_\a^s \right)\left( v_\b-u_\b^s \right)\left( v_\g-u_\g^s \right)}^s$ reads:
\begin{equation}
    \begin{aligned}
        M_{\a\b\g}^s=\frac{1}{\rho^s}\IntV &f^{t}(\bm{r}-\bm{v}\h,\bm{v})\\
        &\cdot\left( v_\a-u_\a^s \right)\left( v_\b-u_\b^s \right)\left( v_\g-u_\g^s \right)
    \end{aligned}
\end{equation}
Using a similar procedure in the calculation of the integral Eq.~\eqref{jy_kin}, we obtain
\begin{equation}\label{M3s}
    \begin{aligned}
        M_{\a\b\g}^s=&\,M_{\a\b\g}^t-\T\h\left( \kr{\a\b}\kr{\g\tau}\right.\\
        &\left.+\kr{\b\g}\kr{\a\tau}+\kr{\g\a}\kr{\b\tau} \right)\pd{\tau}\T+\mathcal{O}\left( \delta^2 \right).
    \end{aligned}
\end{equation}

\emph{Transformation in the collision step.}---The moment after collision $M_{\a\b\g}^{sc}$ can be simply expressed as (see~\cite{CSRD_3D} for a proof):
\begin{equation}
    M_{\a\b\g}^{sc}=\ave{\bk{\left( v_\a^c-u_\a^s \right)\left( v_\b^c-u_\b^s \right)\left( v_\g^c-u_\g^s \right)}^s}.
\end{equation}
Here $v_\a^c$ is the velocity after collision given by Eq.~\eqref{SingleColEq}. Contracting two indexes of $M_{\a\b\g}^{sc}$ gives:
\begin{equation}\label{Mbba1}
    M_{\b\b\a}^{sc}=\ave{\bk{v_\b^c v_\b^c v_\a^c}^s}+\mathcal{O}\left( \delta^2 \right).
\end{equation}
We then substitute Eqs.~\eqref{SingleColEq} and \eqref{Rmat} into Eq.~\eqref{Mbba1}. After derivation, the third-moment $M_{\b\b\a}^{sc}$ after collision is derived:
\begin{equation}\label{Mbba2}
    \begin{aligned}
        M_{\b\b\a}^{sc}=&\,S_{e1}M_{\b\b\a}^{s}+S_{e2}\kr{z\a}M_{\b\b z}^s+S_{e3}M_{zz\a}^s\\
        &+S_{e4}\kr{z\a}M_{zzz}^s+S_{o1}\ep{z\a\tau}M_{\b\b\tau}^s+S_{o2}\ep{z\a\tau}M_{zz\tau}^s,
    \end{aligned}
\end{equation}
where the coefficients $S_{e1\text{--}4}$ and $S_{o1,2}$ are provided in the Appendix~\ref{Appendix} (Eqs.~\eqref{Se1}--\eqref{So2}). In Eq.~\eqref{Mbba2} another third-moment $M_{zz\a}$ appears, requiring us us to derive its transformation in the collision step. Again, through a similar procedure, we have
\begin{equation}\label{Mzza}
    \begin{aligned}
        M_{zz\a}^{sc}=&\,Q_1M_{\b\b\a}^s + Q_2\kr{z\a}M_{\b\b z}^s + Q_3M_{zz\a}^s\\
        &+Q_4\kr{z\a}M_{zzz}^s + Q_5\ep{z\a\tau}M_{\b\b\tau}^s + Q_6\ep{z\a\tau}M_{zz\tau}^s.
    \end{aligned}
\end{equation}
The definitions of coefficients $Q_{1\text{--}6}$ also see Eqs.~\eqref{Q1}--\eqref{Q6} in Appendix~\ref{Appendix}.

\emph{Stationary value of the velocity moment.}---Combining Eqs.~\eqref{M3s}, \eqref{Mbba2}, and \eqref{Mzza} yields:
\begin{equation}\label{TransEqsM3}
    \begin{aligned}
        M_{\b\b\a}^s=&\,M_{\b\b\a}^t-5\T\h\pd{\a}\T,\\
        M_{zz\a}^s=&\,M_{zz\a}^t-\T\h\left( \pd{\a}\T+2\kr{z\a}\pd{z}\T \right),\\
        M_{\b\b\a}^{sc}=&\,S_{e1}M_{\b\b\a}^{s}+S_{e2}\kr{z\a}M_{\b\b z}^s+S_{e3}M_{zz\a}^s\\
        &+S_{e4}\kr{z\a}M_{zzz}^s+S_{o1}\ep{z\a\tau}M_{\b\b\tau}^s+S_{o2}\ep{z\a\tau}M_{zz\tau}^s,\\
        M_{zz\a}^{sc}=&\,Q_1M_{\b\b\a}^s + Q_2\kr{z\a}M_{\b\b z}^s + Q_3M_{zz\a}^s \\
        &+ Q_4\kr{z\a}M_{zzz}^s + Q_5\ep{z\a\tau}M_{\b\b\tau}^s + Q_6\ep{z\a\tau}M_{zz\tau}^s.
    \end{aligned}
\end{equation}
According to Eqs.~\eqref{TransEqsM3}, we obtain the self-consistent equation for $M_{\b\b\a}$ and $M_{zz\a}$ as follows:
\begin{widetext}
    \begin{equation}\label{X_StatEq}
        \begin{aligned}
            \begin{bmatrix}
                S_{e1}-1  & S_{o1}   & S_{e3}  & S_{o2} \\
                -S_{o1}   & S_{e1}-1 & -S_{o2} & S_{e3} \\
                Q_1       & Q_5      & Q_3-1   & Q_6    \\
                -Q_5      & Q_1      & -Q_6    & Q_3-1    \\
            \end{bmatrix}
            \begin{bmatrix}
                M_{\b\b x}\\
                M_{\b\b y}\\
                M_{zzx}\\
                M_{zzy}\\
            \end{bmatrix}
            =\T\h
            \begin{bmatrix}
                S_{e1}  & S_{o1} & S_{e3}  & S_{o2} \\
                -S_{o1} & S_{e1} & -S_{o2} & S_{e3} \\
                Q_1     & Q_5    & Q_3     & Q_6    \\
                -Q_5    & Q_1    & -Q_6    & Q_3    \\
            \end{bmatrix}
            \begin{bmatrix}
                5\pd{x}\T\\
                5\pd{y}\T\\
                \pd{x}\T\\
                \pd{y}\T\\
            \end{bmatrix},
        \end{aligned}
    \end{equation}
    \begin{equation}\label{Y_StatEq}
        \begin{aligned}
            \begin{bmatrix}
                S_{e1}+S_{e2}-1 & S_{e3}+S_{e4} \\
                Q_3+Q_4 & Q_1+Q_2-1\\
            \end{bmatrix}
            \begin{bmatrix}
                M_{\b\b z}\\
                M_{zzz}\\
            \end{bmatrix}
            =\T\h
            \begin{bmatrix}
                S_{e1}+S_{e2} & S_{e3}+S_{e4} \\
                Q_3+Q_4 & Q_1+Q_2\\
            \end{bmatrix}
            \begin{bmatrix}
                5\pd{z}\T\\
                3\pd{z}\T
            \end{bmatrix}.
        \end{aligned}
    \end{equation}

The solution for $M_{\b\b\a}$ is:
\begin{equation}\label{Mbba_form}
    \begin{aligned}
        M_{\b\b\a}=&\,\left( 5\T\h\kr{\a\b} + M^{\perp}_e\krv{\a\b}+M^{\perp}_o\ep{z\a\b} \right.\\
        &\left.+ M^{\parallel}\kr{z\a}\kr{z\b} \right)\pd{\b}\T,\\
        M^{\perp}_e &= \T\h\frac{\left( S_{o2} - 5Q_6 \right)a + \left[ 5\left( Q_3-1 \right) - S_{e3} \right]b}{a^2+b^2}, \\
        M^{\perp}_o &= -\T\h\frac{\left[ 5\left( Q_3-1 \right) - S_{e3} \right]a + \left( 5Q_6-S_{o2} \right)b}{a^2+b^2}, \\
        M^{\parallel} &= \T\h\frac{5\left( Q_1+Q_2-1 \right)-3\left( S_{e3}+S_{e4} \right)}{
            \splitfrac{
                \left( Q_1+Q_2-1 \right)\left( S_{e1}+S_{e2}-1 \right)
            }{
                - \left( Q_3+Q_4 \right)\left( S_{e3}+S_{e4} \right)
            }
        }.
    \end{aligned}
\end{equation}
with $\krv{\a\b}\triangleq\kr{\a\b}-\kr{z\a}\kr{z\b}$ and 
\begin{equation}
    \begin{aligned}
        a&=\left( 1-S_{e1} \right)Q_6 + \left( 1-Q_3 \right)S_{o1} + Q_5S_{e3}+Q_1S_{o2},\\
        b&=\left( 1-Q_3 \right)\left( S_{e1}-1 \right)+Q_1S_{e3}+Q_6S_{o1}-Q_5S_{o2}.
    \end{aligned}
\end{equation}
\end{widetext}

\subsubsection{The kinetic heat flux and thermal conductivity tensor}
Substituting Eq.~\eqref{Mbba_form} into the kinetic energy flux Eq.~\eqref{qKin} yields
\begin{equation}\label{JE_kin}
    J_\a^{\left( E_k \right),\rm{kin}}=\frac{5}{2}\rho\T u_\a+q_\a^{\rm{kin}},
\end{equation}
where the second term here is the kinetic heat flux, given by:
\begin{equation}\label{q_kin}
    \begin{aligned}
        q_\a^{\rm{kin}}=&\,\frac{1}{2}\rho\left( \frac{5}{2}\T\h\kr{\a\b} + M^{\perp}_e\krv{\a\b}\right.\\
        &\left. + M^{\perp}_o\ep{z\a\b} + M^{\parallel}_e\kr{z\a}\kr{z\b} \right)\pd{\b}\T.
    \end{aligned}
\end{equation}
From this, we read off the kinetic thermal conductivity tensor $\kappa_{\a\b}^{\rm{kin}}$ as:
\begin{equation}
    \begin{aligned}
        q_\a^{\rm{kin}}&=-\kappa_{\a\b}^{\rm{kin}}\pd{\b}T,\\
        \kappa_{\a\b}^{\rm{kin}}&=\kappa^{\perp,\rm{kin}}_e\krv{\a\b}+\kappa^{\perp,\rm{kin}}_o\ep{z\a\b}+\kappa^{\parallel,\rm{kin}}\kr{z\a}\kr{z\b}.
    \end{aligned}
\end{equation}
Here $\kappa^{\perp,\rm{kin}}_e$ and $\kappa^{\parallel,\rm{kin}}$ are normal thermal conductivities (even functions of $\theta$), and $\kappa^{\perp,\rm{kin}}_o$ is an odd thermal conductivity (odd function of $\theta$):
\begin{widetext}
\begin{equation}\label{kappa_kin}
    \begin{aligned}
        \kappa^{\perp,\rm{kin}}_e&=-n\frac{k_B^2T}{2m}\h\left\{ \frac{5}{2} + \frac{\left( S_{o2} - 5Q_6 \right)a + \left[ 5\left( Q_3-1 \right) - S_{e3} \right]b}{a^2+b^2} \right\},\\
        \kappa^{\perp,\rm{kin}}_o&=n\frac{k_B^2T}{2m}\h\frac{\left[ 5\left( Q_3-1 \right) - S_{e3} \right]a + \left( 5Q_6-S_{o2} \right)b}{a^2+b^2},\\
        \kappa^{\parallel,\rm{kin}}&=-n\frac{k_B^2T}{2m}\h\left[ \frac{5}{2} + \frac{5\left( Q_1+Q_2-1 \right)-3\left( S_{e3}+S_{e4} \right)}{\left( Q_1+Q_2-1 \right)\left( S_{e1}+S_{e2}-1 \right) - \left( Q_3+Q_4 \right)\left( S_{e3}+S_{e4} \right)} \right],
    \end{aligned}
\end{equation}
\end{widetext}
where $n=\rho/m$ is the particle number density.

\subsection{Heat conduction equation for the 3D-CSRD fluid}
According to Eq.~\eqref{Q}, the general form of conservation equation for energy for the CSRD is
\begin{equation}\label{EnergyEq0}
    \pd{t}\left( \frac{1}{2}\rho\bk{v^2} \right)+\pd{\a}J_\a^{\left( E_k \right)}=0,
\end{equation}
where the first term can be derived using Eq.~\eqref{NS0order} as follows:
\begin{equation}
    \begin{aligned}
        \pd{t}\left( \frac{1}{2}\rho\bk{{v}^2} \right)&=\pd{t}\left( \frac{3}{2}\rho\T+\frac{1}{2}\rho u^2 \right)\\
        &=\pd{t}\left( \frac{3}{2}\rho\T \right)-\rho u_\a\pd{\a}\left( \rho\T \right)+\mathcal{O}\left( \delta^3 \right).
    \end{aligned}
\end{equation}
Using the results of kinetic and collisional energy fluxes Eqs.~\eqref{JE_kin}-\eqref{q_kin} and \eqref{kappa_col}, we obtain the heat conduction equation:
\begin{equation}
    \begin{aligned}
        \pd{t}\left( \frac{3}{2}\rho\T \right)+\pd{\a}\left( \frac{3}{2}\rho\T u_\a \right) = - P\pd{\a}u_\a + \pd{\a}\kappa_{\a\b}\pd{\b}T,
    \end{aligned}
\end{equation}
where $P=\rho\T$ is the pressure and $\kappa_{\a\b}=\kappa_{\a\b}^{\rm{kin}}+\kappa_{\a\b}^{\rm{col}}$ is the thermal conductivity tensor. The total heat flux here is then
\begin{equation}
    q_\a=q_\a^{\rm{kin}}+q_\a^{\rm{col}}=-\kappa_{\a\b}\pd{\b}T.
\end{equation}
According to Eqs.~\eqref{kappa_col} and ~\eqref{kappa_kin}, the thermal conductivity tensor is
\begin{widetext}
    \begin{equation}
        \bm{\kappa}=
        \begin{bmatrix}
            \kappa^{\perp}_e & \kappa^{\perp}_o & 0\\
            -\kappa^{\perp}_o & \kappa^{\perp}_e & 0\\
            0 & 0 & \kappa^{\parallel}\\
        \end{bmatrix}
        =\begin{bmatrix}
            \kappa^{\perp,\rm{kin}}_e + \kappa^{\rm{col}} & \kappa^{\perp,\rm{kin}}_o & 0\\
            -\kappa^{\perp,\rm{kin}}_o & \kappa^{\perp,\rm{kin}}_e + \kappa^{\rm{col}} & 0\\
            0 & 0 & \kappa^{\parallel,\rm{kin}} + \kappa^{\rm{col}}\\
        \end{bmatrix},
    \end{equation}
\end{widetext}
which takes the standard form of thermal conductivity tensor for 3D odd fluids with $C_\infty$ symmetry~\cite{PolyGas2_thCond}.

In this heat conduction equation derived from the kinetic theory, the term related to the mechanical work from the viscous stress is missing. Considering the energy conservation, we modify our result by adding this term phenomenologically:
\begin{equation}\label{ThCoEq}
    \frac{3k_B}{2m}\left[ \pd{t}\left( \rho T \right)+\pd{\a}\left( \rho T u_\a \right) \right] = \sigma_{\a\b}\pd{\b}u_\a + \pd{\a}\kappa_{\a\b}\pd{\b}T.
\end{equation}

\section{Self-diffusion equation for the 3D-CSRD fluid}
To obtain the mass diffusion equation, we sort the fluid particles into two species by tagging some of the particles with $A$ and others with $B$. The density fields for $A$ and $B$ are defined as $\rho_A\left( \bm{r},t \right)$ and $\rho_B\left( \bm{r},t \right)$ respectively. In~\cite{CSRD}, we derived, in a dimension-independent form, the following self-diffusion equation for the density difference $\Delta\rho=\rho_A-\rho_B$ under isothermal and mechanical equilibrium conditions:
\begin{equation}\label{SDeq1}
    \pd{t}\Delta\rho=-\pd{\a}J^D_\a,\qquad J^D_\a=-\pd{\b}\left( D_{\a\b}\Delta\rho \right),
\end{equation}
where $J^D_\a$ is the diffusive mass flux which is the difference between mass fluxes of $A$ and $B$: $J^D_\a=J^A_\a-J^B_\a$, and arises solely from the streaming due to the absence of mass transport in the collision step. Here, the self-diffusion tensor $D_{\a\b}$ takes the form of (see also Eq.~(C19) in~\cite{CSRD}):
\begin{equation}\label{SDT0}
    D_{\a\b} = \frac{\T\h}{2}\left( 2\ave{L}_{\a\b}^{-1} - \kr{\a\b}\right).
\end{equation}
The self-diffusion tensor for the 3D-CSRD is derived straightforwardly by substituting Eqs.~\eqref{SingleColEq} and \eqref{aveR} into Eq.~\eqref{SDT0}:
\begin{equation}
    \bm{D}=\begin{bmatrix}
        D_e^\perp & D_o^\perp & 0 \\
        -D_o^\perp & D_e^\perp & 0 \\
        0 & 0 & D^\parallel \\
    \end{bmatrix},
\end{equation}
where $D^{\perp}_e$ and $D^{\parallel}$ are the normal self-diffusivity, and $D^{\perp}_o$ is the odd self-diffusivity:
\begin{widetext}
    \begin{equation}\label{SDT}
        \begin{aligned}
            D_e^\perp&=\frac{k_BT\h}{2m}\left[ \frac{3\n}{\n-1+e^{-\n}}\frac{3-\left( 1+2\cos\omega \right)\cos\theta}{2\left( 1 - \cos\omega \right)^2 + 3\left( 1+2\cos\omega \right)\left( 1 - \cos\theta \right)}-1 \right] ,\\
            D^\parallel&= \frac{k_BT\h}{2m}\left( \frac{3\n}{\n-1+e^{-\n}}\frac{1}{1-\cos\omega} -1 \right),\\
            D_o^\perp&=-\frac{k_BT\h}{2m}\frac{3\n}{\n-1+e^{-\n}}\frac{\left( 1+2\cos\omega \right)\sin\theta}{2\left( 1 - \cos\omega \right)^2 + 3\left( 1+2\cos\omega \right)\left( 1 - \cos\theta \right)}.
        \end{aligned}
    \end{equation}
\end{widetext}
One can verify that the form of self-diffusivity tensor is consistent with the $C_\infty$ symmetry. Actually, the total density $\rho=\rho_A+\rho_B$ is uniform, which implies a vanishing total mass flux in the system: $J_\a^{(m)}=J^A_\a+J^B_\a=0$. Therefore, the self-diffusion equation Eq.~\eqref{SDeq1} is also expressed, in terms of species A or B, as:
\begin{equation}\label{SelfDiffEq}
    \pd{t}\rho_A=-\pd{\a}J^A_\a,\qquad J^A_\a=-\pd{\b}\left( D_{\a\b}\rho_A \right).
\end{equation}

\begin{figure*}[htbp]
    \centering
    \includegraphics[keepaspectratio, width=1.5\columnwidth]{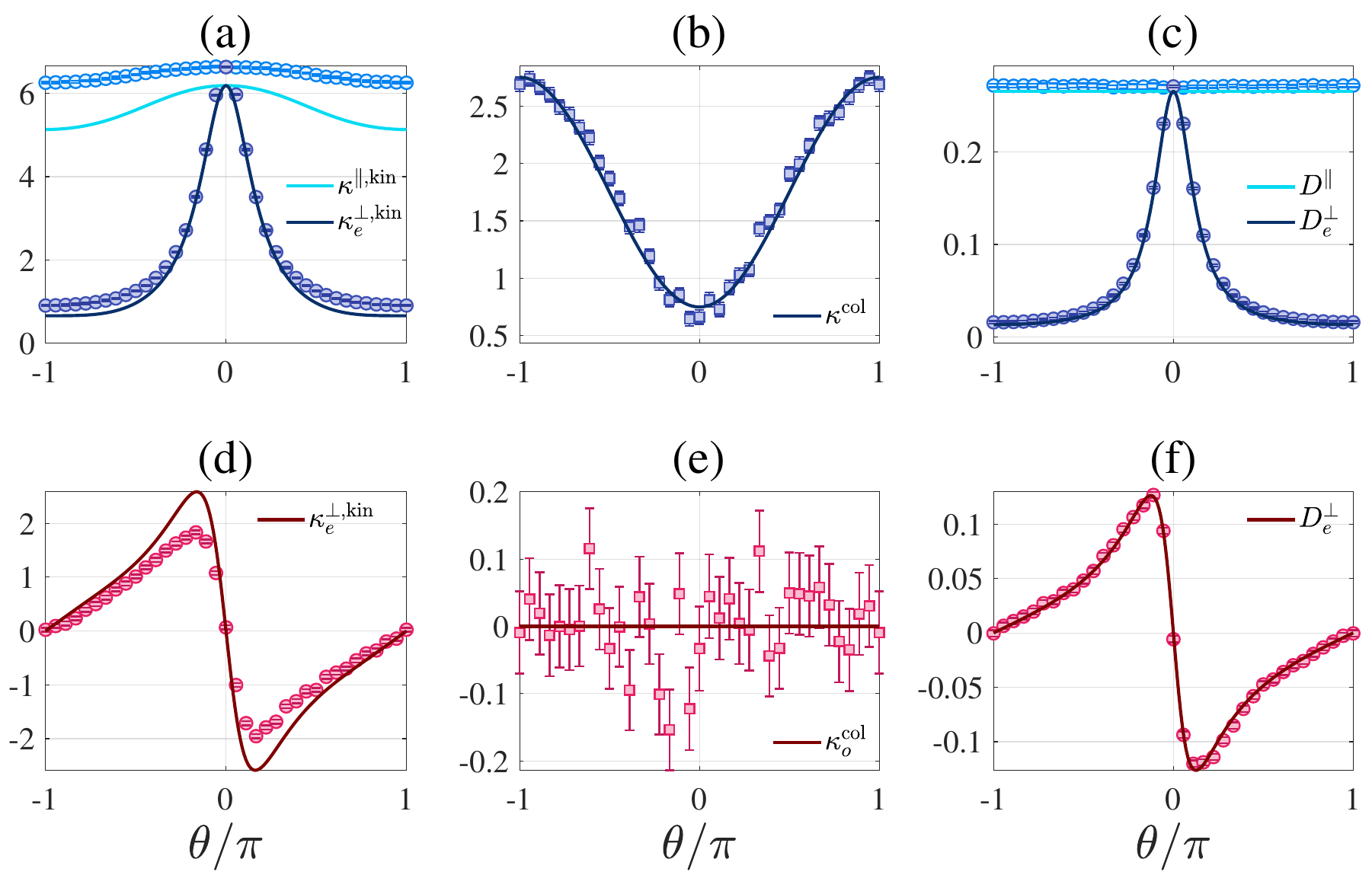}
    \caption{Thermal conductivities (a--b,d--e) and self-diffusivities (c,f) measured in 3D-CSRD simulations with different $\theta$. The normal and odd coefficients are depicted in panels (a--c) and (d--f), respectively. Symbols refer to simulation data and solid lines refer to theoretical results.}
    \label{Fig::Coefficients}
\end{figure*}

\section{Simulation measurements}
In this section, we measure thermal conductivity and self-diffusion tensors of the 3D-CSRD fluid via the nonequilibrium 3D-CSRD simulations.

\subsection{Simulation details}
In simulations, physical quantities are non-dimensionalized by setting $m=1$, $l=1$, and average thermal energy $k_B\ovl{T}=1$. Simulations are performed in a cubic region of size $L=20$ and periodic boundary conditions. We vary $\theta$ and, in a meanwhile, fix $\h=0.1$, $\omega=\pi/3$, $\n=10$ for the measurement of thermal conductivities, and $\n=20$ for the measurement of self-diffusivities.

To determine thermal conductivity or self-diffusivity tensors, we first generate a temperature gradient $\nabla T$ or a density gradient $\nabla\rho_A$ in the system. Then we measure the heat flux $\bm{q}$ or the mass flux of $A$ species, $\bm{J}^A$, in the simulation. Finally, thermal conductivities or the self-diffusivities are determined from the constitutive relations. In our simulations, the $\perp$ components (i.e., $\kappa^{\perp}_e$, $\kappa^{\perp}_o$, $D_e^\perp$, and $D_o^\perp$) and the $\parallel$ components (i.e., $\kappa^{\parallel}$ and $D^\parallel$) are determined by imposing gradients along $y$-axis and $z$-axis, respectively. 

\subsubsection*{Generation of thermal and chemical gradients}
We take the generation of gradients along $y$-axis as an example to illustrate how the temperature/density gradient is created in the simulations. 

In order to impose a temperature gradient $\pd{y}T$, we fixed the temperature of particles in the bottom region $\mathcal{D}_b=\left\{ y|y\in\left[ 0,1 \right] \right\}$ and the middle region $\mathcal{D}_m=\left\{ y|y\in\left[ L/2,L/2+1 \right] \right\}$ at $T_b=0.9$ and $T_m=1.1$, respectively. This temperature control is achieved by applying the Maxwell-Boltzmann scaling thermostat, widely used in the traditional SRD simulations~\cite{MBS_Huang}, in the corresponding region. After a relaxation, the temperature gradient $\pd{y}T\approx 0.02$ is established in the bottom half of the simulation box and an opposite temperature gradient $\pd{y}T\approx -0.02$ is generated in the upper half of the box because of the periodic boundary condition.

The density gradient $\pd{y}\rho_A$ is similarly generated by fixing the density of species $A$ in regions $\mathcal{D}_b$ and $\mathcal{D}_m$ at two different values. Initially, we set $\rho_A=\rho_B=10$. Then after every CSRD step, we convert the species of each particle in $\mathcal{D}_b$ and $\mathcal{D}_m$ to species $A$ with probability $p_b=0.55$ and $p_m=0.45$, respectively. When the system relaxes to the steady state, the density for species $A$ in these two regions becomes ${\rho_A}_b=11$ and ${\rho_A}_m=9$ respectively. Meanwhile, the density gradient $\pd{y}\rho_A\approx -0.2$ is established in the bottom half of the box (and $\pd{y}\rho_A\approx 0.2$ for the upper half).

\begin{figure}[htbp]
    \centering
    \includegraphics[keepaspectratio, width=\columnwidth]{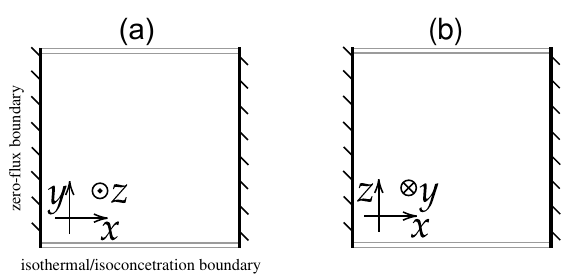}
    \caption{Sketch for the cross-section of the cubic pipe in case study, (a) and (b) depict two different configurations.}\label{Fig::Sketch}
\end{figure}

\begin{figure*}[htbp]
    \centering
    \includegraphics[keepaspectratio, width=1.5\columnwidth]{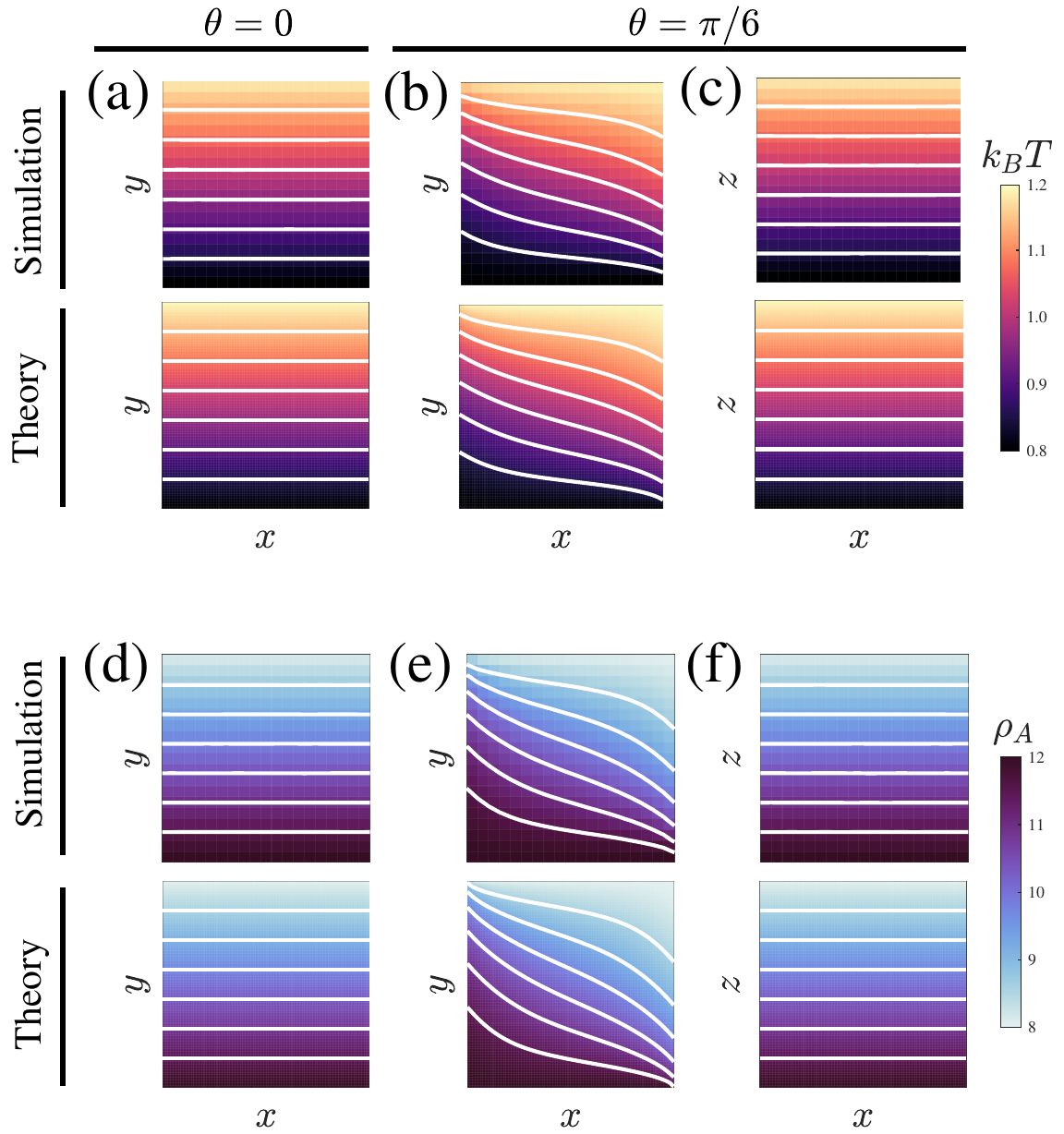}
    \caption{Heat conduction and self-diffusion in a pipe. Cross-sections of the temperature $k_BT$ fields (a--c) and the density $\rho_A$ fields (d--f) under different conditions are shown here (including simulation and theoretical results). The white lines represent the isotherms/isodensity lines. Note that (a,d) are results from the normal fluid ($\theta=0$) and (b--c,e--f) are results from the odd fluid ($\theta=\pi/6$). The zero-flux boundaries are along the $x$-axis and the pipe is along the $z$-axis for (a--b,d--e) and the $y$-axis for (c,f). Parameters: $L=20$, $\h=0.1$, $\omega=\pi/3$, $\n=10$ in (a--c) and $\n=20$ in (b--d), $\left( \phi_b=0.8,\phi_t=1.2 \right)$ in (a--c) and $\left( \phi_b=8,\phi_t=12 \right)$ in (d--f).}
    \label{Fig::CaseStudy}
\end{figure*}

\subsection{Simulation results}
All of the thermal conductivities and self-diffusivities are determined (including kinetic and collision parts) in simulations. We compare these results with the corresponding theoretical predictions in Fig.~\ref{Fig::Coefficients}, displaying that our theoretical results (Eqs.~\eqref{kappa_kin}, \eqref{kappa_col}, and \eqref{SDT}) excellently match simulation measurements. A slight deviation in results for kinetic thermal conductivities in Figs.~\ref{Fig::Coefficients}.(a,d) may be caused by the molecular-chaos hypothesis. It is noted that at $\theta=0$, the model reduces to an isotropic fluid: all odd coefficients vanish, and the $\parallel$ and $\perp$ components for normal transport coefficients become identical. 

\section{Case study}
To further validate the 3D-CSRD model, we study the heat conduction and mass diffusion in a 3D odd fluid confined in a cubic pipe of size $L\times L\times L$. As sketched in Fig.~\ref{Fig::Sketch}, the top and bottom walls are isothermal (for heat conduction) or isoconcentration (for diffusion) boundaries with different temperatures or concentrations, respectively; the left and right walls are zero-flux boundaries; and the front and back boundaries are periodic. The heat conduction/mass diffusion behavior depends on which walls are chosen to align with the $z$-axis (see two different configurations in Fig.~\ref{Fig::Sketch}), owing to the anisotropy of 3D odd fluids. 

For notational convenience, we denote the temperature $T$ and the density $\rho_A$ by a unified variable $\phi$. Under mechanical equilibrium and steady state, the heat conduction equation Eq.~\eqref{ThCoEq} and the self-diffusion equation Eq.~\eqref{SelfDiffEq} reduce to an identical form:
\begin{equation}\label{ctrlEq}
    \left[ \chi_e^\perp\left( \pdu{x}{2}+\pdu{y}{2} \right) + \chi^\parallel\pdu{z}{2} \right]\phi=0,
\end{equation}
with the heat/mass flux
\begin{equation}
    \begin{bmatrix}
        J_x\\
        J_y\\
        J_z
    \end{bmatrix}
    =
    \begin{bmatrix}
        \chi_e^\perp & \chi_o^\perp & 0 \\
        -\chi_o^\perp & \chi_e^\perp & 0 \\
        0 & 0 & \chi^\parallel
    \end{bmatrix}
    \begin{bmatrix}
        \pd{x}\phi\\
        \pd{y}\phi\\
        \pd{z}\phi
    \end{bmatrix},
\end{equation}
where the tensor $\bm{\chi}$ represents the thermal conductivity or self-diffusivity tensor.

The configuration (b) in Fig.~\ref{Fig::Sketch} is trivial, where the odd coefficients are absent in the zero-flux boundary condition, which results in a linear distribution solution of the control equation Eq.~\eqref{ctrlEq} under the given boundary conditions. The simulation results and theory solutions for this example are drawn in Figs.~\ref{Fig::CaseStudy}.(c)(heat conduction) and (f) (mass diffusion), which are the same as the normal fluid case Figs.~\ref{Fig::CaseStudy}.(a,d).

When the pipe is along the $z$-axis [the configuration (a) in Fig.~\ref{Fig::Sketch}], the thermal conduction and mass diffusion behaviors of odd fluids are significantly different from those of normal isotropic fluids. Considering the translation symmetry along $z$-axis, Eq.~\eqref{ctrlEq} becomes
\begin{equation}\label{ctrlEq_z}
    \left( \pdu{x}{2}+\pdu{y}{2} \right)\phi=0
\end{equation}
with the boundary conditions:
\begin{equation}
    \begin{aligned}
        &\left.\phi\right|_{y=0}=\phi_b,\qquad\left.\phi\right|_{y=L}=\phi_t,\\
        &\left.J_x\right|_{x=0,L}=\left.\left( \chi_e^\perp\pd{x}\phi+\chi_o^\perp\pd{y}\phi \right)\right|_{x=0,L}=0.
    \end{aligned}
\end{equation}
This can be numerically solved via the finite difference method. We compare numerical solutions with the corresponding simulation results in Figs.~\ref{Fig::CaseStudy}.(b) and (e) for the heat conduction and the mass diffusion, respectively. The numerical solutions agree well with the simulations. Different from the normal fluid (Figs.~\ref{Fig::CaseStudy}.(a,d)), isotherms and isodensity lines are distorted in the odd fluid because of the transverse flux induced by odd transport coefficients, similar to the case of 2D odd fluid~\cite{CSRD}.

The consistency between the simulation and theoretical results proves that the 3D-CSRD model correctly captures the heat conduction and self-diffusion behaviors of 3D odd fluids.

\section{Conclusion}
In this paper, we derive the heat conduction equation, the self-diffusion equation, and their corresponding constitutive relations for the 3D-CSRD model. These results are verified through simulation measurements and case studies. Combined with the continuity equation and the Navier-Stokes equation derived in our previous work~\cite{CSRD_3D}, this work demonstrates that the 3D-CSRD model can properly describe all hydrodynamic behaviors of 3D odd fluids with $C_\infty$ symmetry. Our work therefore paves the way for large-scale simulation studies of 3D odd fluids.

\section*{Acknowledgment}
This work was supported by the National Natural Science Foundation of China (No. T2325027, No. 12274448).

\appendix
\renewcommand{\theequation}{A\arabic{equation}}
\setcounter{equation}{0}
\section{Coefficients defined in derivations}\label{Appendix}
In derivations of kinetic thermal conductivities, we introduce coefficients $S_{e1\text{--}4}$, $S_{o1,2}$, and $Q_{1\text{--}6}$ for the sake of simplicity. For reference, we summarize these coefficients in this appendix. 

First, we define the following functions of $\omega$ and $\theta$:
\begin{align*}
    &p=\frac{2}{15}\left( 2-\cos\omega-\cos 2\omega \right),\\
    &q=\frac{1}{5}\left( 1+2\cos\omega+2\cos 2\omega \right),\\
    &q_1=\left[ 1-\frac{8}{9}\left( \cos\th-\cos2\th \right) \right]q,\\
    &q_2=\frac{1}{3}\left( 2-\cos\th+2\cos2\th \right)q,\\
    &q_3=-\frac{1}{3}\left( 3-5\cos\th+2\cos2\th \right)q,\\
    &q_4=\frac{1}{3}\left( \sin\th-2\sin2\th \right)q,\\
    &q_5=\frac{2}{105} (4 \cos \omega-3 \cos 2 \omega-3 \cos 3\omega+2) \cos\theta,\\
    &q_6=-\frac{2}{105} (4 \cos \omega-3 \cos 2 \omega-3 \cos 3\omega+2) \sin\theta,\\
    &q_7=\frac{2}{105} (4 \cos \omega-3 \cos2 \omega-3 \cos 3 \omega +2) (3 - \cos \theta),\\
    &q_8=\frac{1}{7} (2 \cos \omega+2 \cos 2 \omega +2 \cos 3 \omega +1) \cos \theta ,\\
    &q_9=-\frac{1}{105} (23 \cos \omega+16 \cos 2 \omega +37 \cos 3 \omega +29) \sin \theta,\\
    &q_{10}=\frac{2}{7} (2 \cos \omega +2 \cos 2 \omega +2 \cos 3 \omega +1) \sin ^2\frac{\theta }{2}.
\end{align*}
The coefficients $S_{e1\text{--}4}$ and $S_{o1,2}$ are:
\begin{equation}\label{Se1}
    \begin{aligned}
        S_{e1}=&\,\ave{\frac{5N-2}{3N^2}}+\ave{\frac{N-1}{3N^2}}q\left( 1+3\cos2\th \right)\\
        &+\ave{\frac{\left( N-1 \right)\left( N-2 \right)}{3N^2}}\left( 1+2\cos\omega \right)\cos\th,
    \end{aligned}
\end{equation}

\begin{equation}\label{Se2}
    \begin{aligned}
        S_{e2}=&\,\ave{\frac{\left( N-1 \right)\left( N-2 \right)}{3N^2}}\left( 1+2\cos\omega \right)\left( 1-\cos\th \right)\\
        &+2\ave{\frac{N-1}{N^2}}q\left( 1-\cos\th \right)\cos\th,
    \end{aligned}
\end{equation}

\begin{equation}\label{Se3}
    \begin{aligned}
        S_{e3}=&\,2\ave{\frac{N-1}{N^2}}q\left( 1-\cos\th \right)\cos\th,
    \end{aligned}
\end{equation}

\begin{equation}\label{Se4}
    \begin{aligned}
        S_{e4}=&\,2\ave{\frac{N-1}{N^2}}q\left( 1-\cos\th \right)^2,
    \end{aligned}
\end{equation}

\begin{equation}\label{So1}
    \begin{aligned}
        S_{o1}=&\,-\ave{\frac{\left( N-1 \right)\left( N-2 \right)}{3N^2}}\left( 1+2\cos\omega \right)\sin\th\\
        &+\ave{\frac{N-1}{N^2}}q\sin2\th,
    \end{aligned}
\end{equation}

\begin{equation}\label{So2}
    \begin{aligned}
        S_{o2}=&\,-2\ave{\frac{N-1}{N^2}}q\left( 1-\cos\th \right)\sin\th.
    \end{aligned}
\end{equation}

The coefficients $Q_{1\text{--}6}$ are:
\begin{equation}\label{Q1}
    \begin{aligned}
        Q_1=&\,\ave{\frac{N-1}{N^2}}p + \ave{\frac{\left( N-1 \right)\left( N-2 \right)}{N^2}}q_5,
    \end{aligned}
\end{equation}

\begin{equation}\label{Q2}
    \begin{aligned}
        Q_2=&\,2\ave{\frac{N-1}{N^2}}p + \ave{\frac{\left( N-1 \right)\left( N-2 \right)}{N^2}}q_7,
    \end{aligned}
\end{equation}

\begin{equation}\label{Q3}
    \begin{aligned}
        Q_3=&\,\ave{\frac{1}{N^2}} + \ave{\frac{N-1}{N^2}}\left( q_1 + 2q_2 \right) \\
        &+ \ave{\frac{\left( N-1 \right)\left( N-2 \right)}{N^2}}q_8,
    \end{aligned}
\end{equation}

\begin{equation}\label{Q4}
    \begin{aligned}
        Q_4=&\,2\ave{\frac{N-1}{N^2}}q_3 + \ave{\frac{\left( N-1 \right)\left( N-2 \right)}{N^2}}q_{10},
    \end{aligned}
\end{equation}

\begin{equation}\label{Q5}
    \begin{aligned}
        Q_5=&\,2\ave{\frac{N-1}{N^2}}q_4 + \ave{\frac{\left( N-1 \right)\left( N-2 \right)}{N^2}}q_{6},
    \end{aligned}
\end{equation}

\begin{equation}\label{Q6}
    \begin{aligned}
        Q_6=&\,\ave{\frac{\left( N-1 \right)\left( N-2 \right)}{N^2}}q_9.
    \end{aligned}
\end{equation}
In the large $\n$ limit, we can simply replace $N$ in expectations in Eqs.~\eqref{Se1}--\eqref{Q6} by $\n$.

\bibliographystyle{apsrev}
\bibliography{CSRD-3D-blx}

\end{document}